\documentclass[%
 reprint,
superscriptaddress,
longbibliography,
groupedaddress,
 amsmath,amssymb,
 aps,
prb,
]{revtex4-1}
\setcitestyle{numbers,square}

\usepackage{braket}
\usepackage{graphicx}
\usepackage{dcolumn}
\usepackage{bm}
\usepackage[normalem]{ulem} 
\usepackage{xcolor}
\usepackage{placeins}
\usepackage{siunitx}
\usepackage{multirow}

\usepackage{hyperref}
\hypersetup{
    colorlinks,%
    citecolor=blue,%
    linkcolor=blue,%
    urlcolor=blue
}

\newcommand{\orcid}[1]{\href{https://orcid.org/#1}{\includegraphics[width=8pt]{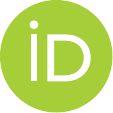}}}

\begin{document}

\title{Substrate suppression of oxidation process in pnictogen monolayers}

\author{Rafael L. H. Freire\orcid{0000-0002-4738-3120}}
\email{rafael.freire@lnnano.cnpem.br}
\affiliation{Ilum School of Science, CNPEM, 13083-970, Campinas, SP, Brazil}

\author{F. Crasto de Lima\orcid{0000-0002-2937-2620}} 
\email{felipe.lima@ilum.cnpem.br}
\affiliation{Ilum School of Science, CNPEM, 13083-970, Campinas, SP, Brazil}

\author{A. Fazzio\orcid{0000-0001-5384-7676}}
\email{adalberto.fazzio@ilum.cnpem.br}
\affiliation{Ilum School of Science, CNPEM, 13083-970, Campinas, SP, Brazil}

\date{\today}

\begin{abstract}


2D materials present an interesting platform for device designs. However, oxidation can drastically change the system's properties, which need to be accounted for. Through {\it ab initio} calculations, we investigated freestanding and SiC-supported As, Sb, and Bi mono-elemental layers. The oxidation process occurs through an O$_2$ spin-state transition, accounted for within the Landau-Zener transition. Additionally, we have investigated the oxidation barriers and the role of spin-orbit coupling. Our calculations pointed out that the presence of SiC substrate reduces the oxidation time scale compared to a freestanding monolayer. We have extracted the energy barrier transition, compatible with our spin-transition analysis. Besides, spin-orbit coupling is relevant to the oxidation mechanisms and alters time scales. The energy barriers decrease as the pnictogen changes from As to Sb to Bi for the freestanding systems, while for SiC-supported, they increase across the pnictogen family. Our computed energy barriers confirm the enhanced robustness against oxidation for the SiC-supported systems.  

\end{abstract}

\maketitle

The realization of two-dimensional (2D) materials  through diverse experimental techniques have increased interest in their technological applications on electronic devices. Particularly, the arising topological insulating phase in bismuthene \cite{SCIENCEreis2017}, antimonene \cite{PRBdominguez2018}, strained arsenene \cite{JPDwang2016, SRrahman2019}, with the former robust against disorder \cite{2DMfocassio2021, PRMpezo2021}, leading to low-power spintronics \cite{COMMPHYSgilbert2021}. However, the experimental conditions towards scalable production of these materials pose great challenges due to their relatively low stability \cite{Alvarez-Quiceno_025025_2017}, mainly at room temperature and in the presence of air (oxygen). Freestanding monoelemental materials, like phosphorene, were shown to be very unstable upon O$_2$-exposure being degraded within a few hours \cite{Ziletti_046801_2015}. Indeed, freestanding monolayer pnictogens (P and As) are more prone to oxidation than other 2D materials presenting the same atomic structure \cite{AMIguo2017}, while the presence of a substrate can alter the oxidation process \cite{Alvarez-Quiceno_025025_2017}.

The O$_2$ molecule occurs naturally in a triplet ($^3\Sigma_g^-$) ground state. On the other hand, under experimental conditions (e.g., photoexcitation \cite{Chan_086403_2003}), O$_2$ molecule can be found in excited singlet states, namely $^1\Delta_g$ and $^1\Sigma_g^+$. The singlet states are more reactive than the ground state triplet, being of great importance in oxidation process \cite{Orellana_155901_2001}. Experimental results over oxidation of 3D-stacked pnictogen systems (down to a few layers), show the robustness of oxidation for the internal layers, while the surface presents oxygen groups \cite{NATCOMMji2016, AMares2016}. Ruled by the higher interlayer bond of heavier pnictogens (compared with the phosphorene), the formation of surface oxide-layer protects the internal layers from oxidation \cite{2DMATassebban2020, CSRzhang2018, AMpumera2017}. There are studies about oxidation on 2D pnictogen materials, however focusing on the freestanding configuration \cite{Kistanov_015010_2017, Kistanov_4308_2018, JMCCkistanov2019, Kistanov_575_2019, Khadiullin_10928_2019, Khadiullin_983_2020}, while not taking into account fundamental aspects, such as the role of triplet-singlet transitions, and spin-orbit effects.

{At the same time, the realization of supported materials through molecular beam epitaxy (MBE) has attracted attention,} for example, Sb/Ag(111) \cite{Shao_2133_2018}, Bi/SiC(0001) \cite{Reis_287_2017} and As/SiC(0001) \cite{JPCCokazaki2023} with a planar structure \cite{Yuan_081003_2020}. Particularly, the topological insulating phase of bismuthene and other pnictogens was predicted when supported on SiC substrate \cite{SCIENCEreis2017, PRBdominguez2018}. While the presence of a substrate can alter the oxidation kinetics of 2D systems \cite{JPCLzhang2016}. In this sense, understanding the mechanisms behind oxygen interaction with those substrate-supported materials is a key point for future experimental {investigations upon applications and routes to improve their stability. }

In this paper, we show that the oxidation process of pnictogen monolayers is considerably lower (slower) when deposited on top of SiC substrate. Taking an {\it ab initio} approach based on the density functional theory (DFT) we investigated the rate of formation of reactive oxygen species, i.e. O$_2$ triplet-singlet transition, close to the materials' surface in the buckled free-standing (FS) form and in the flat structure when on top of SiC substrate (SiC). We connected such rate of formation with the reaction barrier calculated within the nudge elastic band (NEB) method. The FS case {reacts barrierless} with the singlet O$_2$ molecule, while the supported {one presents} a {non-negligible} barrier. Additionally, the barriers found for the triplet O$_2$ molecule are considerably larger for the heavier pnictogen Bi. Our results draw attention to the possible atmospheric stability of supported pnictogens {monolayer}.

\begin{figure}
\includegraphics[width=0.45\textwidth]{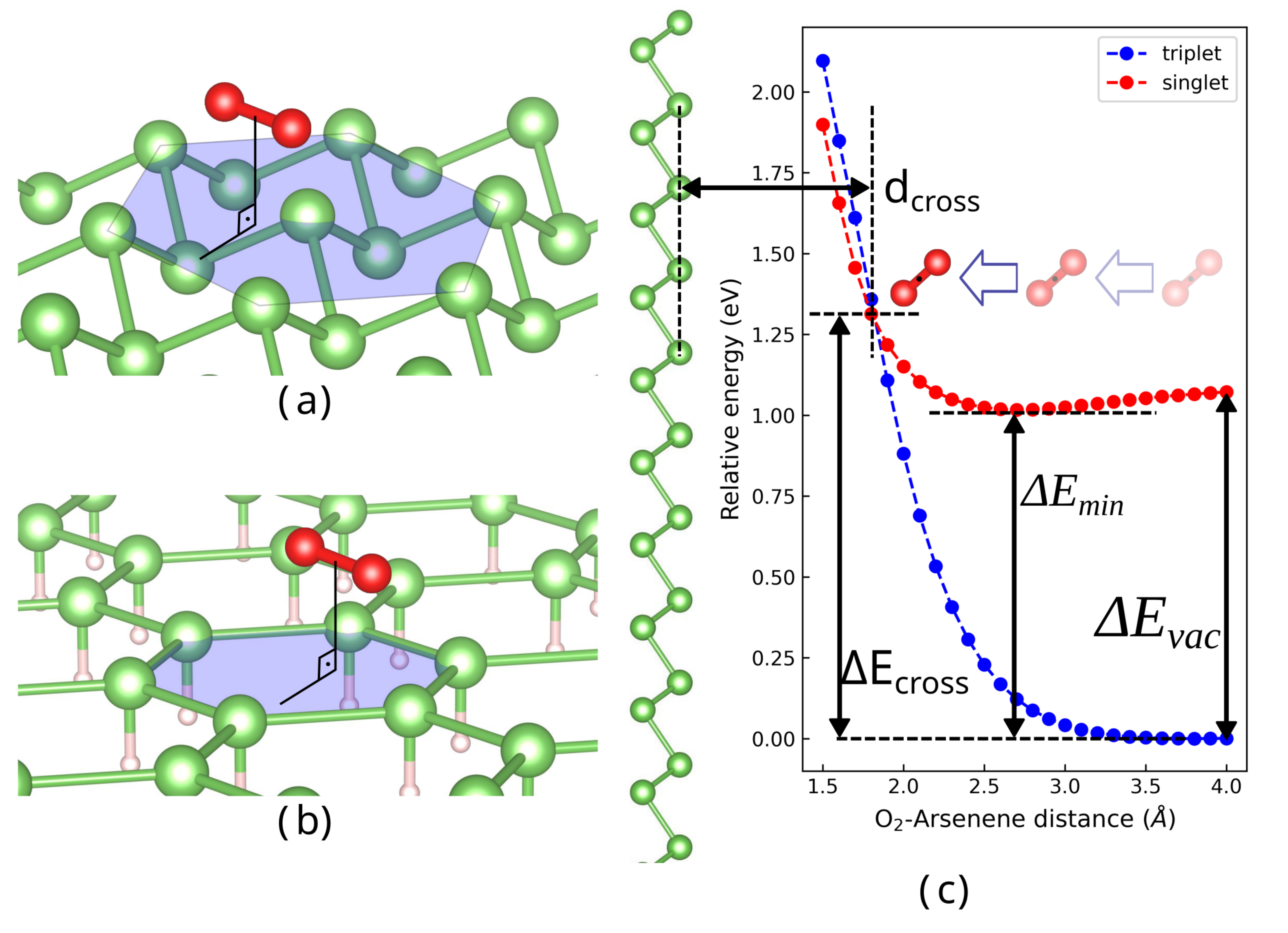}
\caption{\label{fig:O2ads} O$_2$ adsorption model for (a) freestanding and (b) SiC-supported structures, and (c) an example for evaluating the Landau-Zener probabilities, including a few definitions like the distance of the molecule center-of-mass from the 2D material surface at the triplet-singlet crossing ($d_{\rm cross}$), the singlet-triplet energy difference far from the surface ($\Delta E_{vac}$), at the energy minimum ($\Delta E_{min}$).}
\end{figure}

Group-5A elemental monolayers were investigated through spin-polarized calculations based on density functional theory (DFT) \cite{Hohenberg_B864_1964, Kohn_A1133_1965}, performed within the semi-local exchange-correlation functional proposed by Perdew--Burke--Ernzerhof \cite{Perdew_3865_1996}. For the total energies, the electron-ion interactions were considered within the projector augmented wave (PAW) method \cite{PRBblochl1994, PRBkresse1999}, as implemented in the  vienna {\it ab-initio} simulation package (VASP) \cite{PRBkresse1993, PRBkresse1996}. For all calculations, the cutoff energy for the plane-wave expansion of the Kohn--Sham orbitals was set to $400$\,eV, under an energy convergence parameter of $10^{-6}$\,eV, with all atoms relaxed until the atomic forces on every atom were smaller than $10^{-2}$\,eV\,{\AA}$^{-1}$. We considered $3 \times 3$ unit cells with $13$\,{\AA} and $16$\,{\AA} distance between periodic images for FS and SiC-supported systems respectively. A uniform $4\times4\times1$ k-point mesh was considered for the Brillouin zone (BZ) integration.

The {oxidation process} of pnictogen 2D allotropes is known in the literature to be an exothermic process. We calculate the adsorption energy ($E_a$) of a single oxygen atom on the pnictogen surface in its {buckled} freestanding geometry (FS) and in the flat geometry presented when supported on silicon-carbide (SiC-supported) [Fig.~\ref{fig:O2ads}(a) and (b)]. It is worth pointing out that the bismuthene and antimonene on top of SiC form honeycomb lattices, while arsenene has a lower energy triangular lattice \cite{JPCCokazaki2023}, which is considered here. In Table\,\ref{table:Oads},we present our calculations for the adsorption energy
\begin{equation}
    E_a = E_{\rm X+O} - E_{\rm X} - \frac{1}{2} E_{\rm O_2},
\end{equation}
where $E_{X}$ is the pristine pnictogen configuration, $E_{\rm X+O}$ the pnictogen with single oxygen adsorbed on its surface, and $E_{\rm O_2}$ the isolated $O_2$ molecule total energy. Indeed, the adsorption process is still exothermic even for the substrate-supported case. To obtain those adsorption energies we have considered different adsorption sites according to the surface geometry. Thus, in the FS case, we probed on-top, bridge, valley, and hollow sites, while for SiC were on-top, bridge, and hollow sites. For all cases in the lower energy configuration, the oxygen atom forms a bridge between adjacent pnictogen atoms. Comparing the FS with the supported SiC system, we see higher adsorption energies for Sb and Bi, while a decrease is observed for As. Here, the supported As system has a larger tensile strain than the Sb and Bi, when compared to their freestanding structure \cite{JPCCokazaki2023}. The oxygen adsorption, bridging two adjacent As atoms, contributes to lowering the tensile strain, therefore leading to lower adsorption energy.

\begin{figure*}
    \centering
    \includegraphics[width=0.90\textwidth]{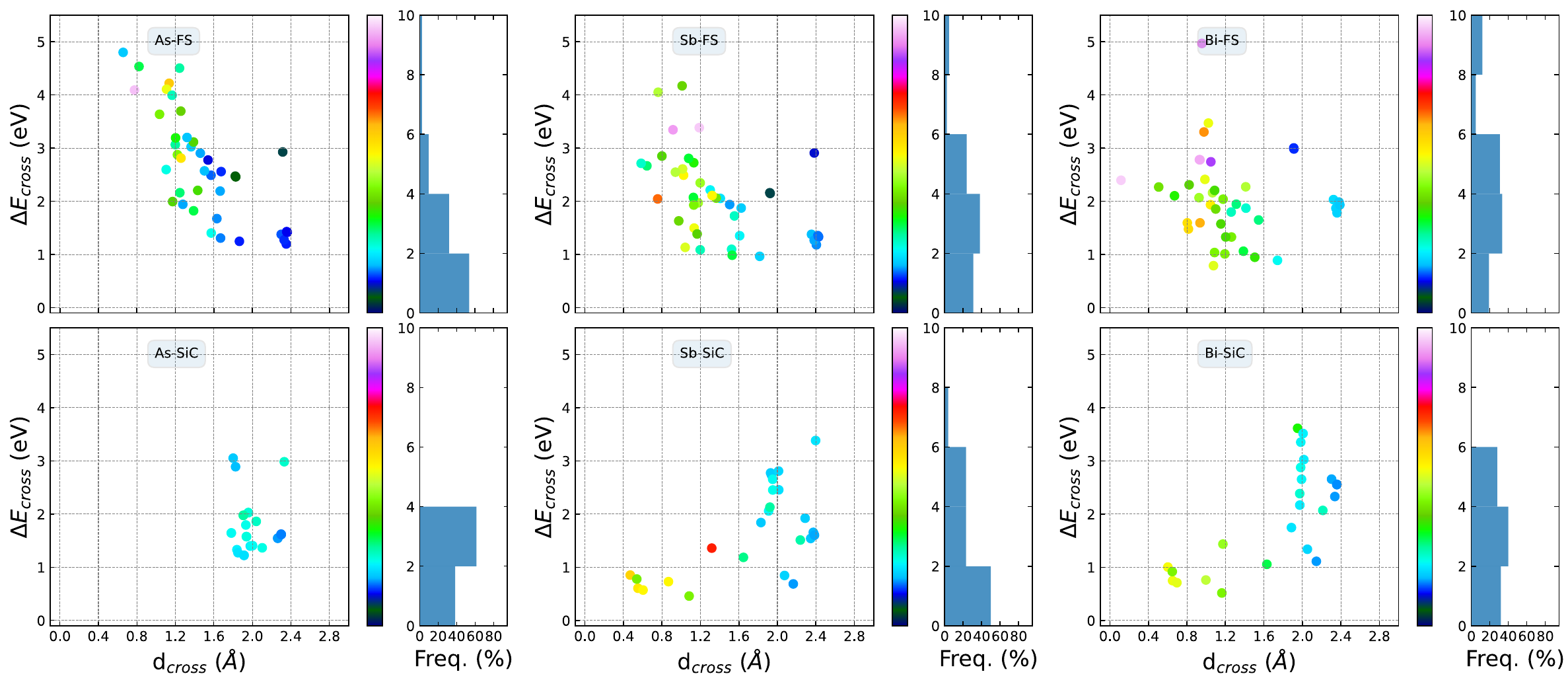}
    \caption{\label{fig:LZmap} Triplet-singlet Landau-Zener transition probabilities ($P_{ts}$) for free-standing (top) and SiC-supported (bottom) systems for a few different adsorption sites, depicted with the triplet-singlet crossing distance ($d_{cross}$) and crossing energy ($\Delta E_{cross}$). $P_{ts}$ is indicated by the color bar, and the histogram indicates the $P_{ts}$ value distribution.}
\end{figure*}

Although there is an indication of a higher exothermic process for As, oxidation can have different reaction time scales for each system. Here we will (i) explore the Landau-Zener probability of transition between oxygen molecule triplet and its most reactive form, oxygen-singlet, close to the pnictogen surfaces and (ii) explore energy barriers for the oxidation process considering the role of the spin-orbit coupling through the nudge elastic band (NEB) method.

\begin{table}
\begin{ruledtabular}
\caption{\label{table:Oads} Oxygen adsorption energy, $E_a$ (eV/O-atom), on pnictogen surfaces on their freestanding (FS) configuration and on silicon carbide-supported (SiC) configuration. The most stable configuration is an epoxy-like bridge bond inclined towards the hexagonal center.}
    \begin{tabular}{cccc}
         phases & As & Sb & Bi  \\
         \hline
         FS        &  \num{-1.01}  &  \num{-1.32}   & \num{-1.06}  \\
         SiC       &  \num{-2.69}  &  \num{-1.15}   & \num{-0.45}  
    \end{tabular}
\end{ruledtabular}
\end{table}



Analyzing the total energy of an O$_2$ molecule close to a materials interface, we see a dependence between the singlet and triplet spin configurations total energies and the molecule distance from the pnictogen surface, as show in Fig.~\ref{fig:O2ads} (c). Away from the surface the singlet and triplet state are separated in energy by $\Delta E_{\rm vac} \sim 1$\,eV, while close to the pnictogen surfaces they present an energy crossing. This crossing implies a transition probability between the two spin states of O$_2$ molecule. Based on the slope of the triplet and singlet curves we have obtained the triplet-singlet transition probabilities ($P_{ts}$) by employing the Landau-Zener relation ($P_{LZ}$) \cite{Zener_696_1932, Harvey_331_2007, Alvarez-Quiceno_025025_2017}
\begin{equation}
    P_{ts} = (1 - P_{LZ})(1 + P_{LZ}),  
\end{equation}
where 
\begin{equation}
P_{LZ} = \exp \left(-\frac{V^2}{hv|F_t - F_s|}\right).
\end{equation}
Here, $V$ is the spin-orbit matrix element of O$_2$ molecule (122 cm$^{-1}$), $v$ the velocity of O$_2$ molecule at room temperature (483.59 m s$^{-1}$), and $F_{i}$ the forces acting on the O$_2$ molecule for each spin state (triplet and singlet) \cite{Alvarez-Quiceno_025025_2017}. It is worth noting that $F_i$ will depend on the materials local adsorption site, and the arriving geometry of the O$_2$ molecule. That is, a single adsorption site can not capture the variations on triplet-single transition, as at experimental conditions this should run over a large distribution of possible sites and molecule geometries (orientation with respect to the surface). Our analysis includes different adsorption sites for both FS and SiC-supported structures and different molecule geometries. This will generate one-dimensional curves as that presented as an example in Fig.~\ref{fig:O2ads}~(c), in which the singlet and triplet potential energy surfaces cross at some point ($d_{cross}$) \cite{Harvey_331_2007}. We extracted information about the (i) triplet-singlet crossing distance ($d_{cross}$); (ii) crossing point relative energy ($\Delta E_{cross}$), (iii) the singlet minimum relative energy ($\Delta E_{min}$), and (iv) the triplet-singlet transition probability ($P_{ts}$).

In Fig.~\ref{fig:LZmap}, we present the triplet-singlet transition probability ($P_{ts}$, in the color bar), mapping it with respect to the crossing relative energies ($\Delta E_{cross}$) and the distance from the surface at the crossing point ($d_{cross}$). In the right panel close to the color bar we represent the $P_{ts}$ statistical distribution. Here, $50\%$ of the FS configurations presented $P_{ts}\,'s < 5\%$, while $60\%$ for SiC-supported. Additionally, the SiC-supported has a probability transition more concentrated around the $2\%$, while the FS configurations present values spreading to higher probabilities. That is, we have a statistical indication that the triplet-single transition is more probable in FS than in the SiC-supported pnictogens.  

In Table~\ref{table:LZstat}, we summarize the average values and mean deviation for the different configurations probed. Despite the significant mean deviation values, we can see that the $P_{ts}$ average for FS is larger than that for SiC-supported, indicating FS as more prone to $O_2$ triplet-singlet transition than SiC-supported, thus facilitating the oxidation process. The crossing distance between the triplet-singlet curves is higher for the SiC-supported than in FS, given the buckled nature of the latter. We see a monotonic growth of $P_{ts}$ when going from As$\rightarrow$Sb$\rightarrow$Bi in the FS case, which is not observed for the SiC system. Furthermore, we see a correlation between $d_{cross}$ with the $P_{ts}$, where the closer to the surface, the larger $P_{ts}$, that is, the surface orbitals interaction with the molecule is ruling the transition. In fact, because of the different bonding nature within the two structures, their orbitals will have different spreading into the vacuum region. In the FS structure, there is a hybridization between in-plane and out-of-plane orbitals forming a $sp^3$ ($s, p_x, p_y, p_z$) bonding, while in the flat SiC-supported, the absence of hybridization between in-plane and out-of-plane orbitals leads to the formation of a $sp^2$ bonding, and a remaining out-of-plane orbital ($p_z$) \cite{Yuan_081003_2020}. Because the $p_z$ orbital is not hybridized in the latter it can possibly spread into larger distances within the vacuum region if compared to the FS structure. Thus, the molecule will feel the presence of the SiC-supported structure at larger distances as a result of the interaction with this out-of-plane orbital depending on the surface site and geometry it approaches. The singlet configuration presents minimum energy when close to the system surface, being the singlet minimum relative energy ($\Delta E_{min}$) lower for the SiC system. This singlet minimum energy is due to unstable physisorbed configurations of the O$_2$ that arise only when constraining the system in the singlet state. As we will show below, such configuration presents a barrierless transition to oxidation and cannot be stabilized on FS systems.

\begin{table}
\begin{ruledtabular}
\caption{\label{table:LZstat} Average values of $\Delta E_{min}$ (eV),  $d_{cross}$ ({\AA}) [shown in Fig.~\ref{fig:O2ads}], and Landau-Zener triplet-singlet probability transition $P_{ts}$ ({\%}) for all configurations tested [Fig.~\ref{fig:LZmap}]. Numbers in parentheses are the standard deviation for the respective quantity.}
\begin{tabular}{ccccc}
\multicolumn{2}{c}{phases} & $\Delta E_{min}$ & $d_{cross}^{O_2^{CM}}$ & $P_{ts}$ \\
         \cline{1-2}\cline{3-5}
\multirow{2}{*}{As} & FS &  1.04 (0.02) & 1.51 (0.46) & 2.46 (1.78) \\
                    &SiC &  0.94 (0.11) & 1.87 (0.28) & 2.15 (1.25) \\
\multirow{2}{*}{Sb} & FS &  0.93 (0.07) & 1.41 (0.54) & 3.33 (2.03) \\
                    &SiC &  0.72 (0.15) & 1.74 (0.64) & 2.85 (1.67) \\
\multirow{2}{*}{Bi} & FS &  0.77 (0.09) & 1.30 (0.55) & 4.25 (2.31) \\
                    &SiC &  0.70 (0.10) & 1.74 (0.61) & 2.77 (1.34) \\        
\end{tabular}
\end{ruledtabular}
\end{table}

\begin{figure}
    \includegraphics[width=\columnwidth]{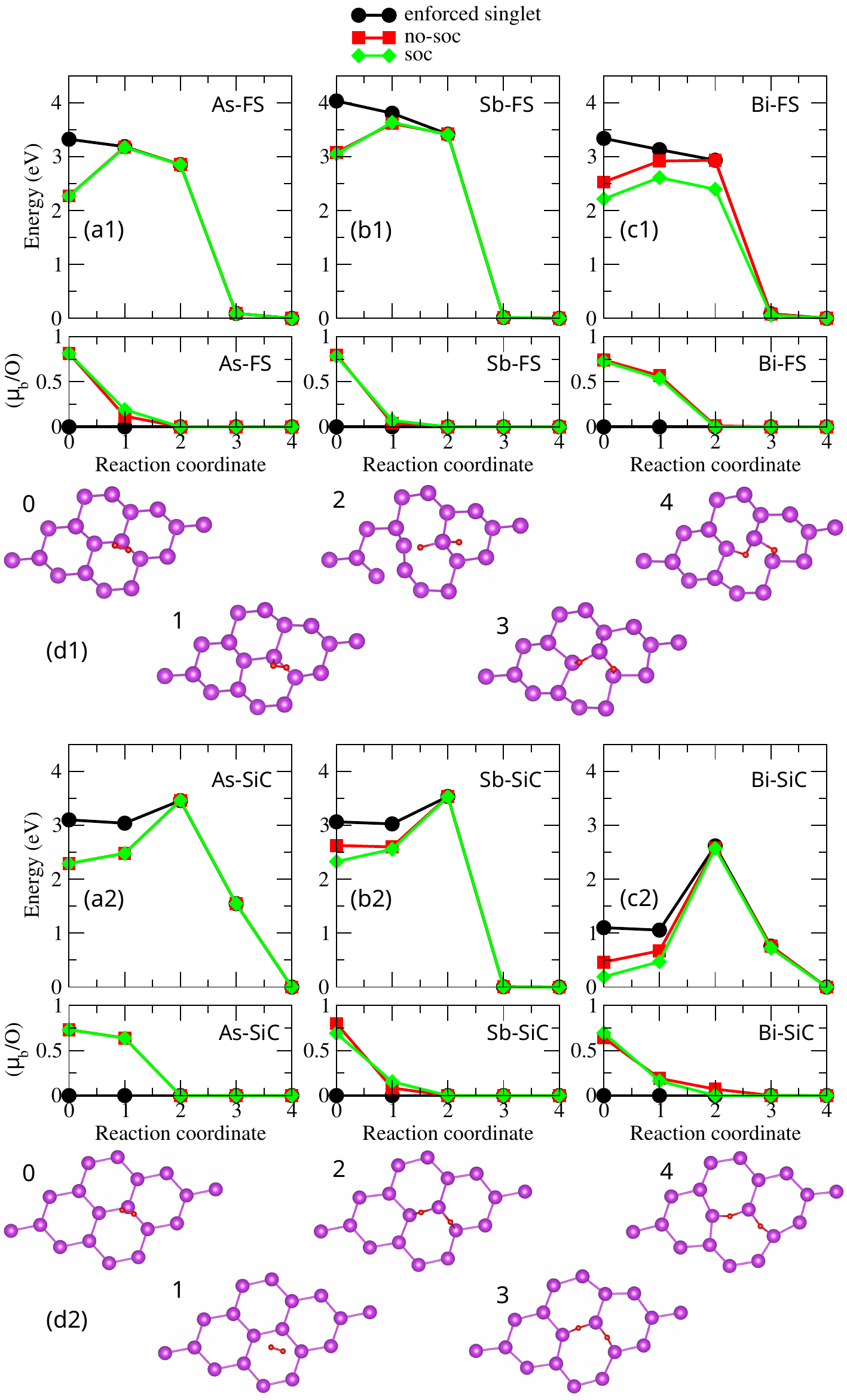}
    \caption{\label{fig:neb} O$_2$ reaction barriers (upper panels) and magnetization along the barrier (lower panels) calculated by the nudge elastic band method, for (a1)-(d1) the FS configuration and (a2)-(d2) the SiC configuration. The atoms' trajectory shown is for the Bi systems, similar geometries are observed for the other systems.}
\end{figure}

Given the scenario for triplet-singlet transition, the reaction rate is also dependent on the energy barrier for both configurations to adsorb on the pnictogen surface. Here we have calculated the energy barrier through the nudge elastic band (NEB) method, considering three scenarios: (i) O$_2$ in an enforced singlet configuration, (ii) O$_2$ with a free spin degree of freedom without spin-orbit coupling, and (iii) a fully relativistic case taking spin-orbit coupling into account. Our results are presented in Fig.~\ref{fig:neb} and summarized in Table~\ref{table:bar}. First, analyzing the enforced singlet case the O$_2$ molecule finds no energy barrier to dissociate over the FS material surface, while for SiC-supported systems there always exists an energy barrier. The singlet energy barrier for the latter is lower for the As and Sb system ($0.36$ and $0.47$\,eV respectively) while a higher value of $1.52$\,eV was found for Bi. We see a different scenario when considering a free spin degree of freedom, here far from the surface the O$_2$ is in a triplet state while through the barrier it changes to a singlet state before dissociation (see the magnetization in the lower panels of Fig.~\ref{fig:neb}). Such behavior is present with or without the spin-orbit effect. This spin transition before the dissociation is dictated by a spin selection rule given the non-magnetic character of oxidized pnictogens \cite{WCMSharvey2014, PRLbehler2005}. The spin-orbit effect is negligible for As and Sb systems, while presenting different effects on Bi. For Bi-FS the spin-orbit coupling lowers both the barrier maximum and the initial state energies, while for Bi-SiC it lowers the initial state keeping the barrier maximum energy. In the singlet states s=0, the spin-orbit contribution vanishes ($\vec{L} \cdot \vec{s}$), while on the triplet state it presents a non-vanishing contribution. For the Bi-FS the triplet state persists higher on the barrier which gives this barrier lowering, while on the Bi-SiC in the barrier maximum, the s=0 state is already defined. 

We see different behavior of the barrier for FS and SiC configuration across the pnictogen group. While for FS, heavier pnictogen present a lower barrier, for supported system the opposite is observed. The decrease in barrier towards heavier pnictogens in FS configuration was also previously observed \cite{Kistanov_575_2019}. For FS system, the Bi system presents a lower energy barrier. Indeed, our Landau-Zener transition probability analysis has shown that the triplet-singlet transition for Bismuth is more favorable than Sb and As. As indicated by the magnetization panels of Fig.~\ref{fig:neb}, the barrier height in the non-strained FS system is ruled by the triplet-singlet transition. On the other hand, the SiC-supported pnictogens are under strain, which can change their interaction {energy} with O$_2$. Bismuth has the largest atomic radius among the pnictogens studied, being under lower strain followed by Sb and As for the SiC supported structure \cite{JPCCokazaki2023}. Such lower strain energy makes the initial configuration (before O$_2$ reaction) lower in energy compared with other pnictogens, leading to a higher barrier for the reaction.

         

\begin{table}
\begin{ruledtabular}
\caption{\label{table:bar} Barrier energies $E_{bar}$ (eV) for O$_2$ reaction on pnictogen surfaces for the initial state in enforced singlet ($s=0$), and triplet ($s=1$) without/with SOC (no-soc/soc).}
    \begin{tabular}{ccccccc}
        system &  FS$^{s=1}_{soc}$ & FS$^{s=1}_{no-soc}$ & FS$^{s=0}$ &  SiC$^{s=1}_{soc}$ & SiC$^{s=1}_{no-soc}$ & SiC$^{s=0}$ \\
         \hline
         As        & 0.90 & 0.91 & 0.00 & 1.17 & 1.17 & 0.36 \\
         Sb        & 0.59 & 0.59 & 0.00 & 1.20 & 0.91 & 0.47 \\
         Bi        & 0.40 & 0.40 & 0.00 & 2.38 & 2.11 & 1.16 \\
    \end{tabular}
\end{ruledtabular}
\end{table}




\begin{figure*}
    \includegraphics[width=1.65\columnwidth]{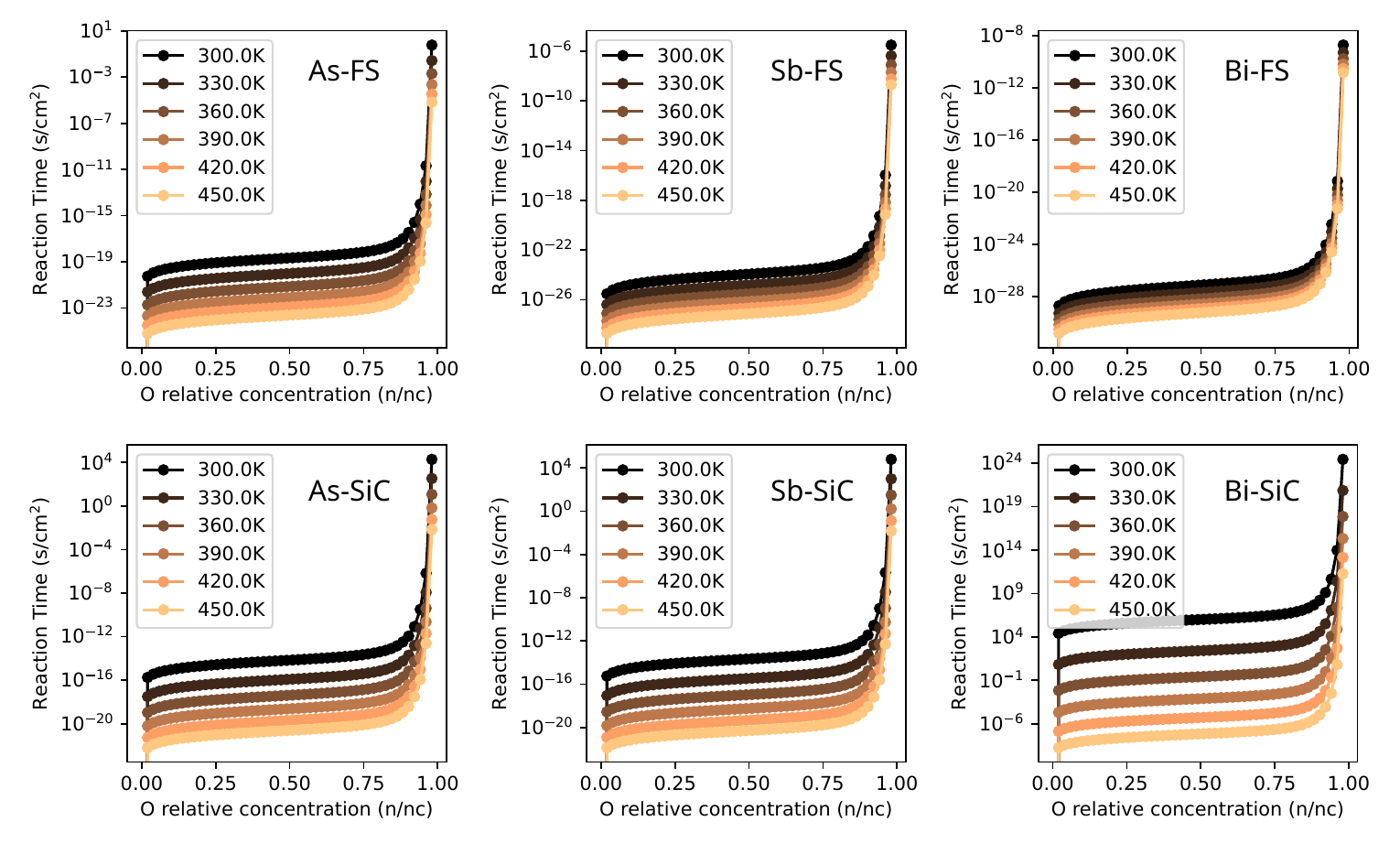}
    \caption{\label{fig:rec-time} Reaction time for oxidation from zero up to oxygen concentration of X$_2$O$_3$ phase (X=As, Sb, Bi), namely critical concentration $n_c$. Upper panels are for FS configuration and lower panels for SiC-supported.}
\end{figure*}

The rate of oxidation for the pristine pnictogen systems can be estimated as
\begin{equation}
    f_0 = \nu e^{\left(-E_b/kT\right)} 
\end{equation}
with $\nu$ the attempt frequency, $E_b$ the calculated barrier energy. In the kinetic theory of gases, for one atmospheric pressure, at 300\,K ($k T = 0.026$\,eV), the number of $O_2$ molecules arriving at a surface per unity of area, per unity of time is
\begin{equation}
    \frac{n}{4}\sqrt{\frac{8kT}{\pi m}} \sim \frac{1.87 \cdot 10^{24}}{{\rm s\cdot cm}^2},
\end{equation}
with $n= 5.1\cdot10^{24}$\,m$^{-3}$ the number of $O_2$ molecules in air per volume at atmospheric pressure/temperature, and $m = 4.9\cdot10^{-26}$\,kg the $O_2$ mass, at $kT = 4.16\cdot10^{-18}$\,kg\,m$^2$s$^{-2}$. 

Such rate of oxidation $f_0$, is valid for the pristine non-oxidized surface. When the system approaches its most stable oxide phase X$_2$O$_3$ (with X=As, Sb, Bi), such rate should vanish. Therefore the rate of oxidation should decay with the surface oxygen concentration $\eta$ from $f_0$ to zero at the critical concentration $\eta_c$ equivalent to the oxygen density in the X$_2$O$_3$ phase. 
\begin{equation}
    f(\eta) = f_0 e^{- \frac{\eta}{\eta_c - \eta}}.
\end{equation}
Given such {oxidation rate}, the reaction time needed for the system to oxidize one cm$^2$ from oxygen concentration zero up to $\eta$ is 
\begin{equation}
    T = 2 \int_0^{\eta} [f(x)]^{-1} dx.
\end{equation}

In Fig.~\ref{fig:rec-time}, we display the reaction time as a function of the relative O concentration $\eta/\eta_c$, for different temperatures.  Here we can see a fast oxidation process for the FS systems. Indeed, experimental results on multilayer pnictogen systems have shown a fast oxidation process on the exposed surface layer \cite{NATCOMMji2016, AMares2016, 2DMATassebban2020}. However, on supported SiC systems, the time scale increases by several orders of magnitude. For As and Sb systems, {despite} the increased time scale, the oxidation process still hinders experimental realization of Arsenene/Antimonene at atmospheric conditions. On the other hand, supported Bi present an oxidation process slow enough to allow an exposition of its surface on atmospheric condition. Increasing the temperature can lead to an oxidation reaction time drastically reduced. For instance, temperatures about $390$\,K should be enough for the supported Bi system to lose its oxidation robustness.

In summary, we have shown the triplet-singlet spin-transition of O$_2$ molecule, rules the oxidation process in monolayer pnictogens. Through our Landau-Zener statistical analysis, we have shown that the FS systems present higher spin-transition probabilities than the SiC-supported ones. By exploring the minimum energy path through the O$_2$ dissociation, we have extracted the barrier transition energy, compatible with our spin-transition analysis. Besides, spin-orbit coupling plays an important role in the oxidation mechanisms and time scales. Particularly, it has a significant effect on SiC-supported systems. The energy barrier presents an inverse dependence with the heavier pnictogen for the FS system (lower for Bi), while a direct dependence is observed for the SiC-supported systems (higher for Bi). The computed barriers confirm the enhanced robustness against oxidation for the SiC-supported systems. Based on that, we have established that according to the reaction time scale for complete oxidation (at 300 \,K), SiC-supported Bi is robust against atmospheric conditions. Our results open a path to explore the optimal 2D systems/substrate interplay aiming {their} experimental {manipulation for further} applications at atmospheric conditions.

\FloatBarrier
\begin{acknowledgments}

The authors acknowledge financial support from the Brazilian agencies FAPESP (grants 20/14067-3, 19/20857-0, and 17/02317-2), CNPq, INCT-Nanocarbono, INCT-Materials Informatics, and Laborat\'{o}rio Nacional de Computa\c{c}\~{a}o Cient\'{i}fica for computer time (project ScafMat2 and emt2D).

\end{acknowledgments}


\bibliography{bib}

\begin{thebibliography}{39}%
\makeatletter
\providecommand \@ifxundefined [1]{%
 \@ifx{#1\undefined}
}%
\providecommand \@ifnum [1]{%
 \ifnum #1\expandafter \@firstoftwo
 \else \expandafter \@secondoftwo
 \fi
}%
\providecommand \@ifx [1]{%
 \ifx #1\expandafter \@firstoftwo
 \else \expandafter \@secondoftwo
 \fi
}%
\providecommand \natexlab [1]{#1}%
\providecommand \enquote  [1]{``#1''}%
\providecommand \bibnamefont  [1]{#1}%
\providecommand \bibfnamefont [1]{#1}%
\providecommand \citenamefont [1]{#1}%
\providecommand \href@noop [0]{\@secondoftwo}%
\providecommand \href [0]{\begingroup \@sanitize@url \@href}%
\providecommand \@href[1]{\@@startlink{#1}\@@href}%
\providecommand \@@href[1]{\endgroup#1\@@endlink}%
\providecommand \@sanitize@url [0]{\catcode `\\12\catcode `\$12\catcode
  `\&12\catcode `\#12\catcode `\^12\catcode `\_12\catcode `\%12\relax}%
\providecommand \@@startlink[1]{}%
\providecommand \@@endlink[0]{}%
\providecommand \url  [0]{\begingroup\@sanitize@url \@url }%
\providecommand \@url [1]{\endgroup\@href {#1}{\urlprefix }}%
\providecommand \urlprefix  [0]{URL }%
\providecommand \Eprint [0]{\href }%
\providecommand \doibase [0]{http://dx.doi.org/}%
\providecommand \selectlanguage [0]{\@gobble}%
\providecommand \bibinfo  [0]{\@secondoftwo}%
\providecommand \bibfield  [0]{\@secondoftwo}%
\providecommand \translation [1]{[#1]}%
\providecommand \BibitemOpen [0]{}%
\providecommand \bibitemStop [0]{}%
\providecommand \bibitemNoStop [0]{.\EOS\space}%
\providecommand \EOS [0]{\spacefactor3000\relax}%
\providecommand \BibitemShut  [1]{\csname bibitem#1\endcsname}%
\let\auto@bib@innerbib\@empty
\bibitem [{\citenamefont {Reis}\ \emph
  {et~al.}(2017{\natexlab{a}})\citenamefont {Reis}, \citenamefont {Li},
  \citenamefont {Dudy}, \citenamefont {Bauernfeind}, \citenamefont {Glass},
  \citenamefont {Hanke}, \citenamefont {Thomale}, \citenamefont {Sch{\"a}fer},\
  and\ \citenamefont {Claessen}}]{SCIENCEreis2017}%
  \BibitemOpen
  \bibfield  {author} {\bibinfo {author} {\bibfnamefont {F.}~\bibnamefont
  {Reis}}, \bibinfo {author} {\bibfnamefont {G.}~\bibnamefont {Li}}, \bibinfo
  {author} {\bibfnamefont {L.}~\bibnamefont {Dudy}}, \bibinfo {author}
  {\bibfnamefont {M.}~\bibnamefont {Bauernfeind}}, \bibinfo {author}
  {\bibfnamefont {S.}~\bibnamefont {Glass}}, \bibinfo {author} {\bibfnamefont
  {W.}~\bibnamefont {Hanke}}, \bibinfo {author} {\bibfnamefont
  {R.}~\bibnamefont {Thomale}}, \bibinfo {author} {\bibfnamefont
  {J.}~\bibnamefont {Sch{\"a}fer}}, \ and\ \bibinfo {author} {\bibfnamefont
  {R.}~\bibnamefont {Claessen}},\ }\bibfield  {title} {\enquote {\bibinfo
  {title} {{Bismuthene on a SiC substrate: A candidate for a high-temperature
  quantum spin Hall material}},}\ }\href {\doibase 10.1126/science.aai8142}
  {\bibfield  {journal} {\bibinfo  {journal} {Science}\ }\textbf {\bibinfo
  {volume} {357}},\ \bibinfo {pages} {287--290} (\bibinfo {year}
  {2017}{\natexlab{a}})}\BibitemShut {NoStop}%
\bibitem [{\citenamefont {Dominguez}\ \emph {et~al.}(2018)\citenamefont
  {Dominguez}, \citenamefont {Scharf}, \citenamefont {Li}, \citenamefont
  {Sch\"afer}, \citenamefont {Claessen}, \citenamefont {Hanke}, \citenamefont
  {Thomale},\ and\ \citenamefont {Hankiewicz}}]{PRBdominguez2018}%
  \BibitemOpen
  \bibfield  {author} {\bibinfo {author} {\bibfnamefont {Fernando}\
  \bibnamefont {Dominguez}}, \bibinfo {author} {\bibfnamefont {Benedikt}\
  \bibnamefont {Scharf}}, \bibinfo {author} {\bibfnamefont {Gang}\ \bibnamefont
  {Li}}, \bibinfo {author} {\bibfnamefont {J\"org}\ \bibnamefont {Sch\"afer}},
  \bibinfo {author} {\bibfnamefont {Ralph}\ \bibnamefont {Claessen}}, \bibinfo
  {author} {\bibfnamefont {Werner}\ \bibnamefont {Hanke}}, \bibinfo {author}
  {\bibfnamefont {Ronny}\ \bibnamefont {Thomale}}, \ and\ \bibinfo {author}
  {\bibfnamefont {Ewelina~M.}\ \bibnamefont {Hankiewicz}},\ }\bibfield  {title}
  {\enquote {\bibinfo {title} {{Testing topological protection of edge states
  in hexagonal quantum spin Hall candidate materials}},}\ }\href {\doibase
  10.1103/PhysRevB.98.161407} {\bibfield  {journal} {\bibinfo  {journal} {Phys.
  Rev. B}\ }\textbf {\bibinfo {volume} {98}},\ \bibinfo {pages} {161407}
  (\bibinfo {year} {2018})}\BibitemShut {NoStop}%
\bibitem [{\citenamefont {ping Wang}\ \emph {et~al.}(2016)\citenamefont {ping
  Wang}, \citenamefont {wen Zhang}, \citenamefont {xiao Ji}, \citenamefont
  {wu~Zhang}, \citenamefont {Li}, \citenamefont {ji~Wang}, \citenamefont {juan
  Ren}, \citenamefont {lian Chen},\ and\ \citenamefont {Yuan}}]{JPDwang2016}%
  \BibitemOpen
  \bibfield  {author} {\bibinfo {author} {\bibfnamefont {Ya}~\bibnamefont {ping
  Wang}}, \bibinfo {author} {\bibfnamefont {Chang}\ \bibnamefont {wen Zhang}},
  \bibinfo {author} {\bibfnamefont {Wei}\ \bibnamefont {xiao Ji}}, \bibinfo
  {author} {\bibfnamefont {Run}\ \bibnamefont {wu~Zhang}}, \bibinfo {author}
  {\bibfnamefont {Ping}\ \bibnamefont {Li}}, \bibinfo {author} {\bibfnamefont
  {Pei}\ \bibnamefont {ji~Wang}}, \bibinfo {author} {\bibfnamefont {Miao}\
  \bibnamefont {juan Ren}}, \bibinfo {author} {\bibfnamefont {Xin}\
  \bibnamefont {lian Chen}}, \ and\ \bibinfo {author} {\bibfnamefont {Min}\
  \bibnamefont {Yuan}},\ }\bibfield  {title} {\enquote {\bibinfo {title}
  {{Tunable quantum spin Hall effect via strain in two-dimensional arsenene
  monolayer}},}\ }\href {\doibase 10.1088/0022-3727/49/5/055305} {\bibfield
  {journal} {\bibinfo  {journal} {Journal of Physics D: Applied Physics}\
  }\textbf {\bibinfo {volume} {49}},\ \bibinfo {pages} {055305} (\bibinfo
  {year} {2016})}\BibitemShut {NoStop}%
\bibitem [{\citenamefont {Rahman}\ \emph {et~al.}(2019)\citenamefont {Rahman},
  \citenamefont {Mahmood},\ and\ \citenamefont
  {Garc\'ia-Su\'arez}}]{SRrahman2019}%
  \BibitemOpen
  \bibfield  {author} {\bibinfo {author} {\bibfnamefont {Gul}\ \bibnamefont
  {Rahman}}, \bibinfo {author} {\bibfnamefont {Asad}\ \bibnamefont {Mahmood}},
  \ and\ \bibinfo {author} {\bibfnamefont {V\'ictor~M.}\ \bibnamefont
  {Garc\'ia-Su\'arez}},\ }\bibfield  {title} {\enquote {\bibinfo {title}
  {{Dynamically Stable Topological Phase of Arsenene}},}\ }\href {\doibase
  10.1038/s41598-019-44444-4} {\bibfield  {journal} {\bibinfo  {journal}
  {Scientific Reports}\ }\textbf {\bibinfo {volume} {9}},\ \bibinfo {pages}
  {7966} (\bibinfo {year} {2019})}\BibitemShut {NoStop}%
\bibitem [{\citenamefont {Focassio}\ \emph {et~al.}(2021)\citenamefont
  {Focassio}, \citenamefont {Schleder}, \citenamefont {Costa}, \citenamefont
  {Fazzio},\ and\ \citenamefont {Lewenkopf}}]{2DMfocassio2021}%
  \BibitemOpen
  \bibfield  {author} {\bibinfo {author} {\bibfnamefont {Bruno}\ \bibnamefont
  {Focassio}}, \bibinfo {author} {\bibfnamefont {Gabriel~R}\ \bibnamefont
  {Schleder}}, \bibinfo {author} {\bibfnamefont {Marcio}\ \bibnamefont
  {Costa}}, \bibinfo {author} {\bibfnamefont {Adalberto}\ \bibnamefont
  {Fazzio}}, \ and\ \bibinfo {author} {\bibfnamefont {Caio}\ \bibnamefont
  {Lewenkopf}},\ }\bibfield  {title} {\enquote {\bibinfo {title} {{Structural
  and electronic properties of realistic two-dimensional amorphous topological
  insulators}},}\ }\href {\doibase 10.1088/2053-1583/abdb97} {\bibfield
  {journal} {\bibinfo  {journal} {2D Materials}\ }\textbf {\bibinfo {volume}
  {8}},\ \bibinfo {pages} {025032} (\bibinfo {year} {2021})}\BibitemShut
  {NoStop}%
\bibitem [{\citenamefont {Pezo}\ \emph {et~al.}(2021)\citenamefont {Pezo},
  \citenamefont {Focassio}, \citenamefont {Schleder}, \citenamefont {Costa},
  \citenamefont {Lewenkopf},\ and\ \citenamefont {Fazzio}}]{PRMpezo2021}%
  \BibitemOpen
  \bibfield  {author} {\bibinfo {author} {\bibfnamefont {Armando}\ \bibnamefont
  {Pezo}}, \bibinfo {author} {\bibfnamefont {Bruno}\ \bibnamefont {Focassio}},
  \bibinfo {author} {\bibfnamefont {Gabriel~R.}\ \bibnamefont {Schleder}},
  \bibinfo {author} {\bibfnamefont {Marcio}\ \bibnamefont {Costa}}, \bibinfo
  {author} {\bibfnamefont {Caio}\ \bibnamefont {Lewenkopf}}, \ and\ \bibinfo
  {author} {\bibfnamefont {Adalberto}\ \bibnamefont {Fazzio}},\ }\bibfield
  {title} {\enquote {\bibinfo {title} {Disorder effects of vacancies on the
  electronic transport properties of realistic topological insulator
  nanoribbons: The case of bismuthene},}\ }\href {\doibase
  10.1103/PhysRevMaterials.5.014204} {\bibfield  {journal} {\bibinfo  {journal}
  {Phys. Rev. Mater.}\ }\textbf {\bibinfo {volume} {5}},\ \bibinfo {pages}
  {014204} (\bibinfo {year} {2021})}\BibitemShut {NoStop}%
\bibitem [{\citenamefont {Gilbert}(2021)}]{COMMPHYSgilbert2021}%
  \BibitemOpen
  \bibfield  {author} {\bibinfo {author} {\bibfnamefont {Matthew~J.}\
  \bibnamefont {Gilbert}},\ }\bibfield  {title} {\enquote {\bibinfo {title}
  {{Topological electronics}},}\ }\href {\doibase 10.1038/s42005-021-00569-5}
  {\bibfield  {journal} {\bibinfo  {journal} {Communications Physics}\ }\textbf
  {\bibinfo {volume} {4}},\ \bibinfo {pages} {70} (\bibinfo {year}
  {2021})}\BibitemShut {NoStop}%
\bibitem [{\citenamefont {Alvarez-Quiceno}\ \emph {et~al.}(2017)\citenamefont
  {Alvarez-Quiceno}, \citenamefont {Miwa}, \citenamefont {Dalpian},\ and\
  \citenamefont {Fazzio}}]{Alvarez-Quiceno_025025_2017}%
  \BibitemOpen
  \bibfield  {author} {\bibinfo {author} {\bibfnamefont {J~C}\ \bibnamefont
  {Alvarez-Quiceno}}, \bibinfo {author} {\bibfnamefont {R~H}\ \bibnamefont
  {Miwa}}, \bibinfo {author} {\bibfnamefont {G~M}\ \bibnamefont {Dalpian}}, \
  and\ \bibinfo {author} {\bibfnamefont {A}~\bibnamefont {Fazzio}},\ }\bibfield
   {title} {\enquote {\bibinfo {title} {{Oxidation of free-standing and
  supported borophene}},}\ }\href {\doibase 10.1088/2053-1583/aa55b6}
  {\bibfield  {journal} {\bibinfo  {journal} {2D Materials}\ }\textbf {\bibinfo
  {volume} {4}},\ \bibinfo {pages} {025025} (\bibinfo {year}
  {2017})}\BibitemShut {NoStop}%
\bibitem [{\citenamefont {Ziletti}\ \emph {et~al.}(2015)\citenamefont
  {Ziletti}, \citenamefont {Carvalho}, \citenamefont {Campbell}, \citenamefont
  {Coker},\ and\ \citenamefont {Castro~Neto}}]{Ziletti_046801_2015}%
  \BibitemOpen
  \bibfield  {author} {\bibinfo {author} {\bibfnamefont {A.}~\bibnamefont
  {Ziletti}}, \bibinfo {author} {\bibfnamefont {A.}~\bibnamefont {Carvalho}},
  \bibinfo {author} {\bibfnamefont {D.~K.}\ \bibnamefont {Campbell}}, \bibinfo
  {author} {\bibfnamefont {D.~F.}\ \bibnamefont {Coker}}, \ and\ \bibinfo
  {author} {\bibfnamefont {A.~H.}\ \bibnamefont {Castro~Neto}},\ }\bibfield
  {title} {\enquote {\bibinfo {title} {{Oxygen Defects in Phosphorene}},}\
  }\href {\doibase 10.1103/PhysRevLett.114.046801} {\bibfield  {journal}
  {\bibinfo  {journal} {Phys. Rev. Lett.}\ }\textbf {\bibinfo {volume} {114}},\
  \bibinfo {pages} {046801} (\bibinfo {year} {2015})}\BibitemShut {NoStop}%
\bibitem [{\citenamefont {Guo}\ \emph {et~al.}(2017)\citenamefont {Guo},
  \citenamefont {Zhou}, \citenamefont {Bai},\ and\ \citenamefont
  {Zhao}}]{AMIguo2017}%
  \BibitemOpen
  \bibfield  {author} {\bibinfo {author} {\bibfnamefont {Yu}~\bibnamefont
  {Guo}}, \bibinfo {author} {\bibfnamefont {Si}~\bibnamefont {Zhou}}, \bibinfo
  {author} {\bibfnamefont {Yizhen}\ \bibnamefont {Bai}}, \ and\ \bibinfo
  {author} {\bibfnamefont {Jijun}\ \bibnamefont {Zhao}},\ }\bibfield  {title}
  {\enquote {\bibinfo {title} {{Oxidation Resistance of Monolayer Group-IV
  Monochalcogenides}},}\ }\href {\doibase 10.1021/acsami.6b16786} {\bibfield
  {journal} {\bibinfo  {journal} {ACS Applied Materials \& Interfaces}\
  }\textbf {\bibinfo {volume} {9}},\ \bibinfo {pages} {12013--12020} (\bibinfo
  {year} {2017})}\BibitemShut {NoStop}%
\bibitem [{\citenamefont {Chan}\ \emph {et~al.}(2003)\citenamefont {Chan},
  \citenamefont {Chen}, \citenamefont {Gong},\ and\ \citenamefont
  {Liu}}]{Chan_086403_2003}%
  \BibitemOpen
  \bibfield  {author} {\bibinfo {author} {\bibfnamefont {Siu-Pang}\
  \bibnamefont {Chan}}, \bibinfo {author} {\bibfnamefont {Gang}\ \bibnamefont
  {Chen}}, \bibinfo {author} {\bibfnamefont {X.~G.}\ \bibnamefont {Gong}}, \
  and\ \bibinfo {author} {\bibfnamefont {Zhi-Feng}\ \bibnamefont {Liu}},\
  }\bibfield  {title} {\enquote {\bibinfo {title} {{Oxidation of Carbon
  Nanotubes by Singlet ${\mathrm{O}}_{2}$}},}\ }\href {\doibase
  10.1103/PhysRevLett.90.086403} {\bibfield  {journal} {\bibinfo  {journal}
  {Phys. Rev. Lett.}\ }\textbf {\bibinfo {volume} {90}},\ \bibinfo {pages}
  {086403} (\bibinfo {year} {2003})}\BibitemShut {NoStop}%
\bibitem [{\citenamefont {Orellana}\ \emph {et~al.}(2001)\citenamefont
  {Orellana}, \citenamefont {da~Silva},\ and\ \citenamefont
  {Fazzio}}]{Orellana_155901_2001}%
  \BibitemOpen
  \bibfield  {author} {\bibinfo {author} {\bibfnamefont {W.}~\bibnamefont
  {Orellana}}, \bibinfo {author} {\bibfnamefont {Ant\^onio J.~R.}\ \bibnamefont
  {da~Silva}}, \ and\ \bibinfo {author} {\bibfnamefont {A.}~\bibnamefont
  {Fazzio}},\ }\bibfield  {title} {\enquote {\bibinfo {title} {{${O}_{2}$
  Diffusion in ${\mathrm{SiO}}_{2}$: Triplet versus Singlet}},}\ }\href
  {\doibase 10.1103/PhysRevLett.87.155901} {\bibfield  {journal} {\bibinfo
  {journal} {Phys. Rev. Lett.}\ }\textbf {\bibinfo {volume} {87}},\ \bibinfo
  {pages} {155901} (\bibinfo {year} {2001})}\BibitemShut {NoStop}%
\bibitem [{\citenamefont {Ji}\ \emph {et~al.}(2016)\citenamefont {Ji},
  \citenamefont {Song}, \citenamefont {Liu}, \citenamefont {Yan}, \citenamefont
  {Huo}, \citenamefont {Zhang}, \citenamefont {Su}, \citenamefont {Liao},
  \citenamefont {Wang}, \citenamefont {Ni}, \citenamefont {Hao},\ and\
  \citenamefont {Zeng}}]{NATCOMMji2016}%
  \BibitemOpen
  \bibfield  {author} {\bibinfo {author} {\bibfnamefont {Jianping}\
  \bibnamefont {Ji}}, \bibinfo {author} {\bibfnamefont {Xiufeng}\ \bibnamefont
  {Song}}, \bibinfo {author} {\bibfnamefont {Jizi}\ \bibnamefont {Liu}},
  \bibinfo {author} {\bibfnamefont {Zhong}\ \bibnamefont {Yan}}, \bibinfo
  {author} {\bibfnamefont {Chengxue}\ \bibnamefont {Huo}}, \bibinfo {author}
  {\bibfnamefont {Shengli}\ \bibnamefont {Zhang}}, \bibinfo {author}
  {\bibfnamefont {Meng}\ \bibnamefont {Su}}, \bibinfo {author} {\bibfnamefont
  {Lei}\ \bibnamefont {Liao}}, \bibinfo {author} {\bibfnamefont {Wenhui}\
  \bibnamefont {Wang}}, \bibinfo {author} {\bibfnamefont {Zhenhua}\
  \bibnamefont {Ni}}, \bibinfo {author} {\bibfnamefont {Yufeng}\ \bibnamefont
  {Hao}}, \ and\ \bibinfo {author} {\bibfnamefont {Haibo}\ \bibnamefont
  {Zeng}},\ }\bibfield  {title} {\enquote {\bibinfo {title} {{Two-dimensional
  antimonene single crystals grown by van der Waals epitaxy}},}\ }\href
  {\doibase 10.1038/ncomms13352} {\bibfield  {journal} {\bibinfo  {journal}
  {Nature Communications}\ }\textbf {\bibinfo {volume} {7}},\ \bibinfo {pages}
  {13352} (\bibinfo {year} {2016})}\BibitemShut {NoStop}%
\bibitem [{\citenamefont {Ares}\ \emph {et~al.}(2016)\citenamefont {Ares},
  \citenamefont {Aguilar-Galindo}, \citenamefont {Rodríguez-San-Miguel},
  \citenamefont {Aldave}, \citenamefont {Díaz-Tendero}, \citenamefont
  {Alcamí}, \citenamefont {Martín}, \citenamefont {Gómez-Herrero},\ and\
  \citenamefont {Zamora}}]{AMares2016}%
  \BibitemOpen
  \bibfield  {author} {\bibinfo {author} {\bibfnamefont {Pablo}\ \bibnamefont
  {Ares}}, \bibinfo {author} {\bibfnamefont {Fernando}\ \bibnamefont
  {Aguilar-Galindo}}, \bibinfo {author} {\bibfnamefont {David}\ \bibnamefont
  {Rodríguez-San-Miguel}}, \bibinfo {author} {\bibfnamefont {Diego~A.}\
  \bibnamefont {Aldave}}, \bibinfo {author} {\bibfnamefont {Sergio}\
  \bibnamefont {Díaz-Tendero}}, \bibinfo {author} {\bibfnamefont {Manuel}\
  \bibnamefont {Alcamí}}, \bibinfo {author} {\bibfnamefont {Fernando}\
  \bibnamefont {Martín}}, \bibinfo {author} {\bibfnamefont {Julio}\
  \bibnamefont {Gómez-Herrero}}, \ and\ \bibinfo {author} {\bibfnamefont
  {Félix}\ \bibnamefont {Zamora}},\ }\bibfield  {title} {\enquote {\bibinfo
  {title} {{Mechanical Isolation of Highly Stable Antimonene under Ambient
  Conditions}},}\ }\href {\doibase https://doi.org/10.1002/adma.201602128}
  {\bibfield  {journal} {\bibinfo  {journal} {Advanced Materials}\ }\textbf
  {\bibinfo {volume} {28}},\ \bibinfo {pages} {6332--6336} (\bibinfo {year}
  {2016})}\BibitemShut {NoStop}%
\bibitem [{\citenamefont {Assebban}\ \emph {et~al.}(2020)\citenamefont
  {Assebban}, \citenamefont {Gibaja}, \citenamefont {Fickert}, \citenamefont
  {Torres}, \citenamefont {Weinreich}, \citenamefont {Wolff}, \citenamefont
  {Gillen}, \citenamefont {Maultzsch}, \citenamefont {Varela}, \citenamefont
  {Rong}, \citenamefont {Loh}, \citenamefont {Michel}, \citenamefont {Zamora},\
  and\ \citenamefont {Abellán}}]{2DMATassebban2020}%
  \BibitemOpen
  \bibfield  {author} {\bibinfo {author} {\bibfnamefont {Mhamed}\ \bibnamefont
  {Assebban}}, \bibinfo {author} {\bibfnamefont {Carlos}\ \bibnamefont
  {Gibaja}}, \bibinfo {author} {\bibfnamefont {Michael}\ \bibnamefont
  {Fickert}}, \bibinfo {author} {\bibfnamefont {Iñigo}\ \bibnamefont
  {Torres}}, \bibinfo {author} {\bibfnamefont {Erik}\ \bibnamefont
  {Weinreich}}, \bibinfo {author} {\bibfnamefont {Stefan}\ \bibnamefont
  {Wolff}}, \bibinfo {author} {\bibfnamefont {Roland}\ \bibnamefont {Gillen}},
  \bibinfo {author} {\bibfnamefont {Janina}\ \bibnamefont {Maultzsch}},
  \bibinfo {author} {\bibfnamefont {Maria}\ \bibnamefont {Varela}}, \bibinfo
  {author} {\bibfnamefont {Sherman Tan~Jun}\ \bibnamefont {Rong}}, \bibinfo
  {author} {\bibfnamefont {Kian~Ping}\ \bibnamefont {Loh}}, \bibinfo {author}
  {\bibfnamefont {Enrique~G}\ \bibnamefont {Michel}}, \bibinfo {author}
  {\bibfnamefont {Félix}\ \bibnamefont {Zamora}}, \ and\ \bibinfo {author}
  {\bibfnamefont {Gonzalo}\ \bibnamefont {Abellán}},\ }\bibfield  {title}
  {\enquote {\bibinfo {title} {{Unveiling the oxidation behavior of
  liquid-phase exfoliated antimony nanosheets}},}\ }\href {\doibase
  10.1088/2053-1583/ab755e} {\bibfield  {journal} {\bibinfo  {journal} {2D
  Materials}\ }\textbf {\bibinfo {volume} {7}},\ \bibinfo {pages} {025039}
  (\bibinfo {year} {2020})}\BibitemShut {NoStop}%
\bibitem [{\citenamefont {Zhang}\ \emph {et~al.}(2018)\citenamefont {Zhang},
  \citenamefont {Guo}, \citenamefont {Chen}, \citenamefont {Wang},
  \citenamefont {Gao}, \citenamefont {Gómez-Herrero}, \citenamefont {Ares},
  \citenamefont {Zamora}, \citenamefont {Zhu},\ and\ \citenamefont
  {Zeng}}]{CSRzhang2018}%
  \BibitemOpen
  \bibfield  {author} {\bibinfo {author} {\bibfnamefont {Shengli}\ \bibnamefont
  {Zhang}}, \bibinfo {author} {\bibfnamefont {Shiying}\ \bibnamefont {Guo}},
  \bibinfo {author} {\bibfnamefont {Zhongfang}\ \bibnamefont {Chen}}, \bibinfo
  {author} {\bibfnamefont {Yeliang}\ \bibnamefont {Wang}}, \bibinfo {author}
  {\bibfnamefont {Hongjun}\ \bibnamefont {Gao}}, \bibinfo {author}
  {\bibfnamefont {Julio}\ \bibnamefont {Gómez-Herrero}}, \bibinfo {author}
  {\bibfnamefont {Pablo}\ \bibnamefont {Ares}}, \bibinfo {author}
  {\bibfnamefont {Félix}\ \bibnamefont {Zamora}}, \bibinfo {author}
  {\bibfnamefont {Zhen}\ \bibnamefont {Zhu}}, \ and\ \bibinfo {author}
  {\bibfnamefont {Haibo}\ \bibnamefont {Zeng}},\ }\bibfield  {title} {\enquote
  {\bibinfo {title} {{Recent progress in 2D group-VA semiconductors: from
  theory to experiment}},}\ }\href {\doibase 10.1039/C7CS00125H} {\bibfield
  {journal} {\bibinfo  {journal} {Chem. Soc. Rev.}\ }\textbf {\bibinfo {volume}
  {47}},\ \bibinfo {pages} {982--1021} (\bibinfo {year} {2018})}\BibitemShut
  {NoStop}%
\bibitem [{\citenamefont {Pumera}\ and\ \citenamefont
  {Sofer}(2017)}]{AMpumera2017}%
  \BibitemOpen
  \bibfield  {author} {\bibinfo {author} {\bibfnamefont {Martin}\ \bibnamefont
  {Pumera}}\ and\ \bibinfo {author} {\bibfnamefont {Zdeněk}\ \bibnamefont
  {Sofer}},\ }\bibfield  {title} {\enquote {\bibinfo {title} {{2D Monoelemental
  Arsenene, Antimonene, and Bismuthene: Beyond Black Phosphorus}},}\ }\href
  {\doibase https://doi.org/10.1002/adma.201605299} {\bibfield  {journal}
  {\bibinfo  {journal} {Advanced Materials}\ }\textbf {\bibinfo {volume}
  {29}},\ \bibinfo {pages} {1605299} (\bibinfo {year} {2017})}\BibitemShut
  {NoStop}%
\bibitem [{\citenamefont {Kistanov}\ \emph {et~al.}(2016)\citenamefont
  {Kistanov}, \citenamefont {Cai}, \citenamefont {Zhou}, \citenamefont
  {Dmitriev},\ and\ \citenamefont {Zhang}}]{Kistanov_015010_2017}%
  \BibitemOpen
  \bibfield  {author} {\bibinfo {author} {\bibfnamefont {Andrey~A}\
  \bibnamefont {Kistanov}}, \bibinfo {author} {\bibfnamefont {Yongqing}\
  \bibnamefont {Cai}}, \bibinfo {author} {\bibfnamefont {Kun}\ \bibnamefont
  {Zhou}}, \bibinfo {author} {\bibfnamefont {Sergey~V}\ \bibnamefont
  {Dmitriev}}, \ and\ \bibinfo {author} {\bibfnamefont {Yong-Wei}\ \bibnamefont
  {Zhang}},\ }\bibfield  {title} {\enquote {\bibinfo {title} {{The role of H2O
  and O2 molecules and phosphorus vacancies in the structure instability of
  phosphorene}},}\ }\href {\doibase 10.1088/2053-1583/4/1/015010} {\bibfield
  {journal} {\bibinfo  {journal} {2D Materials}\ }\textbf {\bibinfo {volume}
  {4}},\ \bibinfo {pages} {015010} (\bibinfo {year} {2016})}\BibitemShut
  {NoStop}%
\bibitem [{\citenamefont {Kistanov}\ \emph {et~al.}(2018)\citenamefont
  {Kistanov}, \citenamefont {Cai}, \citenamefont {Kripalani}, \citenamefont
  {Zhou}, \citenamefont {Dmitriev},\ and\ \citenamefont
  {Zhang}}]{Kistanov_4308_2018}%
  \BibitemOpen
  \bibfield  {author} {\bibinfo {author} {\bibfnamefont {Andrey~A.}\
  \bibnamefont {Kistanov}}, \bibinfo {author} {\bibfnamefont {Yongqing}\
  \bibnamefont {Cai}}, \bibinfo {author} {\bibfnamefont {Devesh~R.}\
  \bibnamefont {Kripalani}}, \bibinfo {author} {\bibfnamefont {Kun}\
  \bibnamefont {Zhou}}, \bibinfo {author} {\bibfnamefont {Sergey~V.}\
  \bibnamefont {Dmitriev}}, \ and\ \bibinfo {author} {\bibfnamefont {Yong-Wei}\
  \bibnamefont {Zhang}},\ }\bibfield  {title} {\enquote {\bibinfo {title} {{A
  first-principles study on the adsorption of small molecules on antimonene:
  oxidation tendency and stability}},}\ }\href {\doibase 10.1039/C8TC00338F}
  {\bibfield  {journal} {\bibinfo  {journal} {J. Mater. Chem. C}\ }\textbf
  {\bibinfo {volume} {6}},\ \bibinfo {pages} {4308--4317} (\bibinfo {year}
  {2018})}\BibitemShut {NoStop}%
\bibitem [{\citenamefont {Kistanov}\ \emph
  {et~al.}(2019{\natexlab{a}})\citenamefont {Kistanov}, \citenamefont
  {Khadiullin}, \citenamefont {Zhou}, \citenamefont {Dmitriev},\ and\
  \citenamefont {Korznikova}}]{JMCCkistanov2019}%
  \BibitemOpen
  \bibfield  {author} {\bibinfo {author} {\bibfnamefont {Andrey~A.}\
  \bibnamefont {Kistanov}}, \bibinfo {author} {\bibfnamefont {Salavat~Kh.}\
  \bibnamefont {Khadiullin}}, \bibinfo {author} {\bibfnamefont {Kun}\
  \bibnamefont {Zhou}}, \bibinfo {author} {\bibfnamefont {Sergey~V.}\
  \bibnamefont {Dmitriev}}, \ and\ \bibinfo {author} {\bibfnamefont {Elena~A.}\
  \bibnamefont {Korznikova}},\ }\bibfield  {title} {\enquote {\bibinfo {title}
  {{Environmental stability of bismuthene: oxidation mechanism and structural
  stability of 2D pnictogens}},}\ }\href {\doibase 10.1039/C9TC03219C}
  {\bibfield  {journal} {\bibinfo  {journal} {J. Mater. Chem. C}\ }\textbf
  {\bibinfo {volume} {7}},\ \bibinfo {pages} {9195--9202} (\bibinfo {year}
  {2019}{\natexlab{a}})}\BibitemShut {NoStop}%
\bibitem [{\citenamefont {Kistanov}\ \emph
  {et~al.}(2019{\natexlab{b}})\citenamefont {Kistanov}, \citenamefont
  {Khadiullin}, \citenamefont {Dmitriev},\ and\ \citenamefont
  {Korznikova}}]{Kistanov_575_2019}%
  \BibitemOpen
  \bibfield  {author} {\bibinfo {author} {\bibfnamefont {Andrey~A.}\
  \bibnamefont {Kistanov}}, \bibinfo {author} {\bibfnamefont {Salavat~Kh.}\
  \bibnamefont {Khadiullin}}, \bibinfo {author} {\bibfnamefont {Sergey~V.}\
  \bibnamefont {Dmitriev}}, \ and\ \bibinfo {author} {\bibfnamefont {Elena~A.}\
  \bibnamefont {Korznikova}},\ }\bibfield  {title} {\enquote {\bibinfo {title}
  {{A First-Principles Study on the Adsorption of Small Molecules on Arsenene:
  Comparison of Oxidation Kinetics in Arsenene, Antimonene, Phosphorene, and
  InSe}},}\ }\href {\doibase https://doi.org/10.1002/cphc.201801070} {\bibfield
   {journal} {\bibinfo  {journal} {ChemPhysChem}\ }\textbf {\bibinfo {volume}
  {20}},\ \bibinfo {pages} {575--580} (\bibinfo {year}
  {2019}{\natexlab{b}})}\BibitemShut {NoStop}%
\bibitem [{\citenamefont {Khadiullin}\ \emph {et~al.}(2019)\citenamefont
  {Khadiullin}, \citenamefont {Kistanov}, \citenamefont {Ustiuzhanina},
  \citenamefont {Davletshin}, \citenamefont {Zhou}, \citenamefont {Dmitriev},\
  and\ \citenamefont {Korznikova}}]{Khadiullin_10928_2019}%
  \BibitemOpen
  \bibfield  {author} {\bibinfo {author} {\bibfnamefont {Salavat~Kh.}\
  \bibnamefont {Khadiullin}}, \bibinfo {author} {\bibfnamefont {Andrey~A.}\
  \bibnamefont {Kistanov}}, \bibinfo {author} {\bibfnamefont {Svetlana~V.}\
  \bibnamefont {Ustiuzhanina}}, \bibinfo {author} {\bibfnamefont {Artur~R.}\
  \bibnamefont {Davletshin}}, \bibinfo {author} {\bibfnamefont {Kun}\
  \bibnamefont {Zhou}}, \bibinfo {author} {\bibfnamefont {Sergey~V.}\
  \bibnamefont {Dmitriev}}, \ and\ \bibinfo {author} {\bibfnamefont {Elena~A.}\
  \bibnamefont {Korznikova}},\ }\bibfield  {title} {\enquote {\bibinfo {title}
  {{First-Principles Study of Interaction of Bismuthene with Small Gas
  Molecules}},}\ }\href {\doibase https://doi.org/10.1002/slct.201903002}
  {\bibfield  {journal} {\bibinfo  {journal} {ChemistrySelect}\ }\textbf
  {\bibinfo {volume} {4}},\ \bibinfo {pages} {10928--10933} (\bibinfo {year}
  {2019})}\BibitemShut {NoStop}%
\bibitem [{\citenamefont {Khadiullin}\ \emph {et~al.}(2020)\citenamefont
  {Khadiullin}, \citenamefont {Davletshin}, \citenamefont {Zhou},\ and\
  \citenamefont {Korznikova}}]{Khadiullin_983_2020}%
  \BibitemOpen
  \bibfield  {author} {\bibinfo {author} {\bibfnamefont {Salavat}\ \bibnamefont
  {Khadiullin}}, \bibinfo {author} {\bibfnamefont {Artur}\ \bibnamefont
  {Davletshin}}, \bibinfo {author} {\bibfnamefont {Kun}\ \bibnamefont {Zhou}},
  \ and\ \bibinfo {author} {\bibfnamefont {Elena}\ \bibnamefont {Korznikova}},\
  }\bibfield  {title} {\enquote {\bibinfo {title} {{Analysis of Chemical
  Activity of Bismuthene in the Presence of Environment Gas Molecules by Means
  of Ab Initio Calculations}},}\ }in\ \href@noop {} {\emph {\bibinfo
  {booktitle} {TMS 2020 149th Annual Meeting {\&} Exhibition Supplemental
  Proceedings}}}\ (\bibinfo  {publisher} {Springer International Publishing},\
  \bibinfo {address} {Cham},\ \bibinfo {year} {2020})\ pp.\ \bibinfo {pages}
  {983--991}\BibitemShut {NoStop}%
\bibitem [{\citenamefont {Shao}\ \emph {et~al.}(2018)\citenamefont {Shao},
  \citenamefont {Liu}, \citenamefont {Cheng}, \citenamefont {Wu}, \citenamefont
  {Liu}, \citenamefont {Liu}, \citenamefont {Wang}, \citenamefont {Zhu},
  \citenamefont {Wang}, \citenamefont {Shi}, \citenamefont {Ibrahim},
  \citenamefont {Sun}, \citenamefont {Wang},\ and\ \citenamefont
  {Gao}}]{Shao_2133_2018}%
  \BibitemOpen
  \bibfield  {author} {\bibinfo {author} {\bibfnamefont {Yan}\ \bibnamefont
  {Shao}}, \bibinfo {author} {\bibfnamefont {Zhong-Liu}\ \bibnamefont {Liu}},
  \bibinfo {author} {\bibfnamefont {Cai}\ \bibnamefont {Cheng}}, \bibinfo
  {author} {\bibfnamefont {Xu}~\bibnamefont {Wu}}, \bibinfo {author}
  {\bibfnamefont {Hang}\ \bibnamefont {Liu}}, \bibinfo {author} {\bibfnamefont
  {Chen}\ \bibnamefont {Liu}}, \bibinfo {author} {\bibfnamefont {Jia-Ou}\
  \bibnamefont {Wang}}, \bibinfo {author} {\bibfnamefont {Shi-Yu}\ \bibnamefont
  {Zhu}}, \bibinfo {author} {\bibfnamefont {Yu-Qi}\ \bibnamefont {Wang}},
  \bibinfo {author} {\bibfnamefont {Dong-Xia}\ \bibnamefont {Shi}}, \bibinfo
  {author} {\bibfnamefont {Kurash}\ \bibnamefont {Ibrahim}}, \bibinfo {author}
  {\bibfnamefont {Jia-Tao}\ \bibnamefont {Sun}}, \bibinfo {author}
  {\bibfnamefont {Ye-Liang}\ \bibnamefont {Wang}}, \ and\ \bibinfo {author}
  {\bibfnamefont {Hong-Jun}\ \bibnamefont {Gao}},\ }\bibfield  {title}
  {\enquote {\bibinfo {title} {{Epitaxial Growth of Flat Antimonene Monolayer:
  A New Honeycomb Analogue of Graphene}},}\ }\href {\doibase
  10.1021/acs.nanolett.8b00429} {\bibfield  {journal} {\bibinfo  {journal}
  {Nano Letters}\ }\textbf {\bibinfo {volume} {18}},\ \bibinfo {pages}
  {2133--2139} (\bibinfo {year} {2018})},\ \bibinfo {note} {pMID:
  29457727}\BibitemShut {NoStop}%
\bibitem [{\citenamefont {Reis}\ \emph
  {et~al.}(2017{\natexlab{b}})\citenamefont {Reis}, \citenamefont {Li},
  \citenamefont {Dudy}, \citenamefont {Bauernfeind}, \citenamefont {Glass},
  \citenamefont {Hanke}, \citenamefont {Thomale}, \citenamefont {Schäfer},\
  and\ \citenamefont {Claessen}}]{Reis_287_2017}%
  \BibitemOpen
  \bibfield  {author} {\bibinfo {author} {\bibfnamefont {F.}~\bibnamefont
  {Reis}}, \bibinfo {author} {\bibfnamefont {G.}~\bibnamefont {Li}}, \bibinfo
  {author} {\bibfnamefont {L.}~\bibnamefont {Dudy}}, \bibinfo {author}
  {\bibfnamefont {M.}~\bibnamefont {Bauernfeind}}, \bibinfo {author}
  {\bibfnamefont {S.}~\bibnamefont {Glass}}, \bibinfo {author} {\bibfnamefont
  {W.}~\bibnamefont {Hanke}}, \bibinfo {author} {\bibfnamefont
  {R.}~\bibnamefont {Thomale}}, \bibinfo {author} {\bibfnamefont
  {J.}~\bibnamefont {Schäfer}}, \ and\ \bibinfo {author} {\bibfnamefont
  {R.}~\bibnamefont {Claessen}},\ }\bibfield  {title} {\enquote {\bibinfo
  {title} {{Bismuthene on a SiC substrate: A candidate for a high-temperature
  quantum spin Hall material}},}\ }\href {\doibase 10.1126/science.aai8142}
  {\bibfield  {journal} {\bibinfo  {journal} {Science}\ }\textbf {\bibinfo
  {volume} {357}},\ \bibinfo {pages} {287--290} (\bibinfo {year}
  {2017}{\natexlab{b}})}\BibitemShut {NoStop}%
\bibitem [{\citenamefont {Okazaki}\ \emph {et~al.}(2023)\citenamefont
  {Okazaki}, \citenamefont {Furlan~de Oliveira}, \citenamefont {Freire},
  \citenamefont {Fazzio},\ and\ \citenamefont {Crasto~de
  Lima}}]{JPCCokazaki2023}%
  \BibitemOpen
  \bibfield  {author} {\bibinfo {author} {\bibfnamefont {Anderson~K.}\
  \bibnamefont {Okazaki}}, \bibinfo {author} {\bibfnamefont {Rafael}\
  \bibnamefont {Furlan~de Oliveira}}, \bibinfo {author} {\bibfnamefont {Rafael
  Luiz~Heleno}\ \bibnamefont {Freire}}, \bibinfo {author} {\bibfnamefont
  {Adalberto}\ \bibnamefont {Fazzio}}, \ and\ \bibinfo {author} {\bibfnamefont
  {Felipe}\ \bibnamefont {Crasto~de Lima}},\ }\bibfield  {title} {\enquote
  {\bibinfo {title} {{Uncovering the Structural Evolution of Arsenene on SiC
  Substrate}},}\ }\href {\doibase 10.1021/acs.jpcc.3c00938} {\bibfield
  {journal} {\bibinfo  {journal} {The Journal of Physical Chemistry C}\
  }\textbf {\bibinfo {volume} {127}},\ \bibinfo {pages} {7894--7899} (\bibinfo
  {year} {2023})}\BibitemShut {NoStop}%
\bibitem [{\citenamefont {Yuan}\ \emph {et~al.}(2020)\citenamefont {Yuan},
  \citenamefont {Zhang}, \citenamefont {Sun}, \citenamefont {Liu},
  \citenamefont {Yao},\ and\ \citenamefont {Wang}}]{Yuan_081003_2020}%
  \BibitemOpen
  \bibfield  {author} {\bibinfo {author} {\bibfnamefont {Peiwen}\ \bibnamefont
  {Yuan}}, \bibinfo {author} {\bibfnamefont {Teng}\ \bibnamefont {Zhang}},
  \bibinfo {author} {\bibfnamefont {Jiatao}\ \bibnamefont {Sun}}, \bibinfo
  {author} {\bibfnamefont {Liwei}\ \bibnamefont {Liu}}, \bibinfo {author}
  {\bibfnamefont {Yugui}\ \bibnamefont {Yao}}, \ and\ \bibinfo {author}
  {\bibfnamefont {Yeliang}\ \bibnamefont {Wang}},\ }\bibfield  {title}
  {\enquote {\bibinfo {title} {{Recent progress in 2D group-V elemental
  monolayers: fabrications and properties}},}\ }\href {\doibase
  10.1088/1674-4926/41/8/081003} {\bibfield  {journal} {\bibinfo  {journal}
  {Journal of Semiconductors}\ }\textbf {\bibinfo {volume} {41}},\ \bibinfo
  {pages} {081003} (\bibinfo {year} {2020})}\BibitemShut {NoStop}%
\bibitem [{\citenamefont {Zhang}\ \emph {et~al.}(2016)\citenamefont {Zhang},
  \citenamefont {Yin}, \citenamefont {Liu}, \citenamefont {Li}, \citenamefont
  {Zhang},\ and\ \citenamefont {Guo}}]{JPCLzhang2016}%
  \BibitemOpen
  \bibfield  {author} {\bibinfo {author} {\bibfnamefont {Zhuhua}\ \bibnamefont
  {Zhang}}, \bibinfo {author} {\bibfnamefont {Jun}\ \bibnamefont {Yin}},
  \bibinfo {author} {\bibfnamefont {Xiaofei}\ \bibnamefont {Liu}}, \bibinfo
  {author} {\bibfnamefont {Jidong}\ \bibnamefont {Li}}, \bibinfo {author}
  {\bibfnamefont {Jiahuan}\ \bibnamefont {Zhang}}, \ and\ \bibinfo {author}
  {\bibfnamefont {Wanlin}\ \bibnamefont {Guo}},\ }\bibfield  {title} {\enquote
  {\bibinfo {title} {{Substrate-Sensitive Graphene Oxidation}},}\ }\href
  {\doibase 10.1021/acs.jpclett.6b00062} {\bibfield  {journal} {\bibinfo
  {journal} {The Journal of Physical Chemistry Letters}\ }\textbf {\bibinfo
  {volume} {7}},\ \bibinfo {pages} {867--873} (\bibinfo {year}
  {2016})}\BibitemShut {NoStop}%
\bibitem [{\citenamefont {Hohenberg}\ and\ \citenamefont
  {Kohn}(1964)}]{Hohenberg_B864_1964}%
  \BibitemOpen
  \bibfield  {author} {\bibinfo {author} {\bibfnamefont {P.}~\bibnamefont
  {Hohenberg}}\ and\ \bibinfo {author} {\bibfnamefont {W.}~\bibnamefont
  {Kohn}},\ }\bibfield  {title} {\enquote {\bibinfo {title} {{Inhomogeneous
  Electron Gas}},}\ }\href {\doibase 10.1103/physrev.136.b864} {\bibfield
  {journal} {\bibinfo  {journal} {Phys. Rev.}\ }\textbf {\bibinfo {volume}
  {136}},\ \bibinfo {pages} {B864--B871} (\bibinfo {year} {1964})}\BibitemShut
  {NoStop}%
\bibitem [{\citenamefont {Kohn}\ and\ \citenamefont
  {Sham}(1965)}]{Kohn_A1133_1965}%
  \BibitemOpen
  \bibfield  {author} {\bibinfo {author} {\bibfnamefont {W.}~\bibnamefont
  {Kohn}}\ and\ \bibinfo {author} {\bibfnamefont {L.~J.}\ \bibnamefont
  {Sham}},\ }\bibfield  {title} {\enquote {\bibinfo {title} {{Self-consistent
  equations including exchange and correlation effects}},}\ }\href {\doibase
  10.1103/physrev.140.a1133} {\bibfield  {journal} {\bibinfo  {journal} {Phys.
  Rev.}\ }\textbf {\bibinfo {volume} {140}},\ \bibinfo {pages} {A1133--A1138}
  (\bibinfo {year} {1965})}\BibitemShut {NoStop}%
\bibitem [{\citenamefont {Perdew}\ \emph {et~al.}(1996)\citenamefont {Perdew},
  \citenamefont {Burke},\ and\ \citenamefont {Ernzerhof}}]{Perdew_3865_1996}%
  \BibitemOpen
  \bibfield  {author} {\bibinfo {author} {\bibfnamefont {J.~P.}\ \bibnamefont
  {Perdew}}, \bibinfo {author} {\bibfnamefont {K.}~\bibnamefont {Burke}}, \
  and\ \bibinfo {author} {\bibfnamefont {M.}~\bibnamefont {Ernzerhof}},\
  }\bibfield  {title} {\enquote {\bibinfo {title} {Generalized gradient
  approximation made simple},}\ }\href {\doibase 10.1103/PhysRevLett.77.3865}
  {\bibfield  {journal} {\bibinfo  {journal} {Phys. Rev. Lett.}\ }\textbf
  {\bibinfo {volume} {77}},\ \bibinfo {pages} {3865--3868} (\bibinfo {year}
  {1996})}\BibitemShut {NoStop}%
\bibitem [{\citenamefont {Bl\"ochl}(1994)}]{PRBblochl1994}%
  \BibitemOpen
  \bibfield  {author} {\bibinfo {author} {\bibfnamefont {P.~E.}\ \bibnamefont
  {Bl\"ochl}},\ }\bibfield  {title} {\enquote {\bibinfo {title} {{Projector
  augmented-wave method}},}\ }\href {\doibase 10.1103/PhysRevB.50.17953}
  {\bibfield  {journal} {\bibinfo  {journal} {Phys. Rev. B}\ }\textbf {\bibinfo
  {volume} {50}},\ \bibinfo {pages} {17953--17979} (\bibinfo {year}
  {1994})}\BibitemShut {NoStop}%
\bibitem [{\citenamefont {Kresse}\ and\ \citenamefont
  {Joubert}(1999)}]{PRBkresse1999}%
  \BibitemOpen
  \bibfield  {author} {\bibinfo {author} {\bibfnamefont {G.}~\bibnamefont
  {Kresse}}\ and\ \bibinfo {author} {\bibfnamefont {D.}~\bibnamefont
  {Joubert}},\ }\bibfield  {title} {\enquote {\bibinfo {title} {{From ultrasoft
  pseudopotentials to the projector augmented-wave method}},}\ }\href {\doibase
  10.1103/PhysRevB.59.1758} {\bibfield  {journal} {\bibinfo  {journal} {Phys.
  Rev. B}\ }\textbf {\bibinfo {volume} {59}},\ \bibinfo {pages} {1758--1775}
  (\bibinfo {year} {1999})}\BibitemShut {NoStop}%
\bibitem [{\citenamefont {Kresse}\ and\ \citenamefont
  {Hafner}(1993)}]{PRBkresse1993}%
  \BibitemOpen
  \bibfield  {author} {\bibinfo {author} {\bibfnamefont {G.}~\bibnamefont
  {Kresse}}\ and\ \bibinfo {author} {\bibfnamefont {J.}~\bibnamefont
  {Hafner}},\ }\bibfield  {title} {\enquote {\bibinfo {title} {{Ab Initio
  Molecular Dynamics for Open-Shell Transition Metals}},}\ }\href {\doibase
  10.1103/PhysRevB.48.13115} {\bibfield  {journal} {\bibinfo  {journal} {Phys.
  Rev. B}\ }\textbf {\bibinfo {volume} {48}},\ \bibinfo {pages} {13115--13126}
  (\bibinfo {year} {1993})}\BibitemShut {NoStop}%
\bibitem [{\citenamefont {Kresse}\ and\ \citenamefont
  {Furthm{\"{u}}ller}(1996)}]{PRBkresse1996}%
  \BibitemOpen
  \bibfield  {author} {\bibinfo {author} {\bibfnamefont {G.}~\bibnamefont
  {Kresse}}\ and\ \bibinfo {author} {\bibfnamefont {J.}~\bibnamefont
  {Furthm{\"{u}}ller}},\ }\bibfield  {title} {\enquote {\bibinfo {title}
  {{Efficient iterative schemes for ab initio total-energy calculations using a
  plane-wave basis set}},}\ }\href {\doibase 10.1103/PhysRevB.54.11169}
  {\bibfield  {journal} {\bibinfo  {journal} {Phys. Rev. B}\ }\textbf {\bibinfo
  {volume} {54}},\ \bibinfo {pages} {11169--11186} (\bibinfo {year}
  {1996})}\BibitemShut {NoStop}%
\bibitem [{\citenamefont {Zener}\ and\ \citenamefont
  {Fowler}(1932)}]{Zener_696_1932}%
  \BibitemOpen
  \bibfield  {author} {\bibinfo {author} {\bibfnamefont {Clarence}\
  \bibnamefont {Zener}}\ and\ \bibinfo {author} {\bibfnamefont {Ralph~Howard}\
  \bibnamefont {Fowler}},\ }\bibfield  {title} {\enquote {\bibinfo {title}
  {Non-adiabatic crossing of energy levels},}\ }\href {\doibase
  10.1098/rspa.1932.0165} {\bibfield  {journal} {\bibinfo  {journal}
  {{Proceedings of the Royal Society of London. Series A, Containing Papers of
  a Mathematical and Physical Character}}\ }\textbf {\bibinfo {volume} {137}},\
  \bibinfo {pages} {696--702} (\bibinfo {year} {1932})}\BibitemShut {NoStop}%
\bibitem [{\citenamefont {Harvey}(2007)}]{Harvey_331_2007}%
  \BibitemOpen
  \bibfield  {author} {\bibinfo {author} {\bibfnamefont {Jeremy~N.}\
  \bibnamefont {Harvey}},\ }\bibfield  {title} {\enquote {\bibinfo {title}
  {{Understanding the kinetics of spin-forbidden chemical reactions}},}\ }\href
  {\doibase 10.1039/B614390C} {\bibfield  {journal} {\bibinfo  {journal} {Phys.
  Chem. Chem. Phys.}\ }\textbf {\bibinfo {volume} {9}},\ \bibinfo {pages}
  {331--343} (\bibinfo {year} {2007})}\BibitemShut {NoStop}%
\bibitem [{\citenamefont {Harvey}(2014)}]{WCMSharvey2014}%
  \BibitemOpen
  \bibfield  {author} {\bibinfo {author} {\bibfnamefont {Jeremy~N.}\
  \bibnamefont {Harvey}},\ }\bibfield  {title} {\enquote {\bibinfo {title}
  {{Spin-forbidden reactions: computational insight into mechanisms and
  kinetics}},}\ }\href {\doibase https://doi.org/10.1002/wcms.1154} {\bibfield
  {journal} {\bibinfo  {journal} {WIREs Computational Molecular Science}\
  }\textbf {\bibinfo {volume} {4}},\ \bibinfo {pages} {1--14} (\bibinfo {year}
  {2014})}\BibitemShut {NoStop}%
\bibitem [{\citenamefont {Behler}\ \emph {et~al.}(2005)\citenamefont {Behler},
  \citenamefont {Delley}, \citenamefont {Lorenz}, \citenamefont {Reuter},\ and\
  \citenamefont {Scheffler}}]{PRLbehler2005}%
  \BibitemOpen
  \bibfield  {author} {\bibinfo {author} {\bibfnamefont {J\"org}\ \bibnamefont
  {Behler}}, \bibinfo {author} {\bibfnamefont {Bernard}\ \bibnamefont
  {Delley}}, \bibinfo {author} {\bibfnamefont {S\"onke}\ \bibnamefont
  {Lorenz}}, \bibinfo {author} {\bibfnamefont {Karsten}\ \bibnamefont
  {Reuter}}, \ and\ \bibinfo {author} {\bibfnamefont {Matthias}\ \bibnamefont
  {Scheffler}},\ }\bibfield  {title} {\enquote {\bibinfo {title} {Dissociation
  of ${\mathrm{o}}_{2}$ at al(111): The role of spin selection rules},}\ }\href
  {\doibase 10.1103/PhysRevLett.94.036104} {\bibfield  {journal} {\bibinfo
  {journal} {Phys. Rev. Lett.}\ }\textbf {\bibinfo {volume} {94}},\ \bibinfo
  {pages} {036104} (\bibinfo {year} {2005})}\BibitemShut {NoStop}%
\end{thebibliography}%

\end{document}